\documentclass[11pt, superscriptaddress]{revtex4}
\usepackage{textcomp}
\usepackage{amsmath}
\newcommand{\RN}[1]{%
  \textup{\uppercase\expandafter{\romannumeral#1}}%
}
\newtheorem{theorem}{Theorem}
\usepackage{amssymb}
\usepackage{graphicx}
\usepackage[utf8]{inputenc}
\usepackage{epstopdf}
\usepackage{float}
\usepackage{hyperref}
\usepackage[a4paper,bindingoffset=0.5in,left=0.5in,right=0.5in,top=1in,bottom=1in,footskip=.25in]{geometry}
\usepackage{color}
\begin{document}
\color{black}
\title{  CARTER CONSTANT AND SUPERINTEGRABILITY } 
\author{K Rajesh Nayak}
\email[E-mail:]{rajesh@iiserkol.ac.in}
\affiliation{ Center of Excellence in Space Sciences, India and Department of Physical Sciences, \\
Indian Institute of Science Education and Research Kolkata, Mohanpur, India - 741246. }
\author{Payel Mukhopadhyay}
\email[E-mail:]{payelmuk@stanford.edu }
\affiliation{ Stanford University, Department of Physics, 382 Via Pueblo Mall, Stanford, CA, 94305}
\affiliation{ Center of Excellence in Space Sciences, India and Department of Physical Sciences, \\
Indian Institute of Science Education and Research Kolkata, Mohanpur, India - 741246. }
\date{December 21, 2017 } 
\begin{abstract}
Carter constant is a non-trivial conserved quantity of motion of a particle moving in stationary axisymmetric spacetime. In the version of the theorem originally given by Carter, due to the presence of two Killing vectors, the system effectively has two degrees of freedom. We propose an extension to the first version of Carter's theorem to a system having three degrees of freedom to find two functionally independent Carter-like integrals of motion. We further  generalize the theorem to a dynamical system with $N$ degrees of freedom. We further study the implications of Carter Constant to Superintegrability and present a different approach to probe a Superintegrable system. Our formalism gives another viewpoint to a Superintegrable system using the simple observation of separable Hamiltonian according to Carter's criteria. We then give some examples by constructing some 2-Dimensional superintegrable systems based on this idea and also show that all 3-D simple classical Superintegrable potentials are also Carter separable.
\end{abstract}
\maketitle

\section{Introduction}

Carter constant is a non-trivial conserved quantity of motion in stationary axisymmetric spacetime 
such as the Kerr solution rendering the equations of motion 
integrable~\cite{Carter:2003abc,C:1973abc,M:1973}. It is a manifestation of one of the hidden 
symmetries of the spacetime. In a general axisymmetric spacetime, there might be no 
independent fourth constant of motion.  It has been shown  by Carter that if the corresponding 
Klein-Gordon equation is separable, then the existence of a fourth constant of motion is 
ensured~\cite{C:1973abc}.
%% In systems, where there is translational symmetry, the Carter constant reduces to the linear momentum and similarly for spherically symmetric systems, the constant is just the total angular momentum\cite{Rosquist:20077abc,Fel:1999abc,Sajal:2015abc}. 
The fourth constant of motion has been extremely useful in studying the geodesics of motion 
of a particle in Kerr spacetime. The Carter's constant along with energy, axial angular
 momentum, and particle rest mass provide the four conserved quantities necessary to 
 integrate  all orbital equations.  The actual physical meaning of Carter constant which 
 appears as a part of the separability  conditions in Hamilton-Jacobi formalism, is still not clearly known. 
 One approach is to look for Newtonian systems that would give rise to non-trivial Carter-like constants.  Refs. \cite{Will:2009abc, Sajal:2015abc} have
given Carter-like constants in Newtonian dynamics.  For example, angular momentum of a system plays an important role in understanding the physics of a rotating system. Because of the interrelation between the components of angular momentum, only two independent scalar quantities can be constructed out of it. Conventionally, they are the azimuthal component of the angular momentum $L_{z}$ and the square of the angular momentum $L^2$. In some well studied scenarios like problems having spherical symmetry, the Carter constant reduces to $L^2$ \cite{Rosquist:20077abc,Fel:1999abc}.  However in general $L^2$ need not be conserved, for example in static or stationary  axisymmetric spacetimes, $L_{z}$ is conserved but $L^2$ is not. In such situations, Carter constant may be used to define $L^2$. 

Carter constant was first given for systems in 4-D  stationary and axially symmetric
spacetime \cite{Carter:2003abc}.  It has been shown by Walker and Penrose that 
for vacuum, Petrov type D solutions admit Carter-like constants \cite{Wa:1970abc}. In 
addition , Ramachandra has  shown that Petrov type D solutions allow a Carter constant even if they are  non-vacuum and asymptotically non-flat ~\cite{Ramachandra:2003abc}.  
One specific case is the Kerr metric and the constant is applicable only to a 
general system having two degrees of freedom ~\cite{C:1973abc,K:1963}.  Another approach for
finding Carter constant is through Killing tensors \cite{Wa:1970abc}, which is a rather brute-force and physically
non-intuitive method. Though computationally hard, in principle it is straightforward to find such 
Killing tensors in higher dimensional spacetimes as well.  \\
 In this short paper, we show that Carter-like constants can also exist in systems with more degrees of freedom. Systems with three degrees  of freedom can exist in 4-D  or higher dimensional spaces. These emerging constants of motion in higher dimensional systems might be useful in revealing several hidden properties of such systems. Closely following the approach given by Carter ~\cite{C:1973abc}, our approach depends on the 
  form and separability of the Hamiltonian and hence can be applied to any Hamiltonian system
  with $N$-degrees of freedom. Furthermore,  the analysis of these constants may be done in general relativistic framework in a variety of different metrics corresponding to different spacetimes or in Newtonian mechanics for a better understanding of Carter constant in general. 

Our primary idea of this work is the following theorem given by  Carter where the  constant of motion can be found out by an inspection of the Hamiltonian \cite{C:1973abc}. 

For a Hamiltonian having the form:
\begin{equation}
H = \frac{1}{2} (\frac{H_r + H_{\mu}}{U_{r} + U_{\mu}})\,,
\label{eq:H2}
\end{equation}
where $U_{r}$, $U_{\mu}$ are single variable functions of coordinates $r$ and ${\mu}$ respectively, where $H_{r}$ is independent of $p_{\mu}$ and of all other coordinate functions other than $r$ and $H_{\mu}$ is independent of $p_{r}$ and of all other coordinate functions other than $p_{\mu}$ then, 
\begin{equation}
\kappa = \frac{U_{r} H_{\mu} - U_{\mu} H_{r}}{U_{r} + U_{\mu}}\,,
\label{eq:K2}
\end{equation}
is a constant of motion. In this  article, we would generalize the theorem by proving that a similar kind of theorem is valid for a system with $N$-degrees of freedom as well.

One of the interesting consequences of integrability of physical systems is the possibility of the existence of Superintegrable systems in nature where roughly by Superintegrability, one implies a syatem having more number of independent integrals of motion than the degrees of freedom available for the system. Some familiar examples, such as the Kepler problem and the harmonic oscillator, have been known since the time of Laplace. Superintegrable systems are extremely important for developing insight into physical principles, for they can be solved algebraically as well as analytically. Such systems are special since they allow maximum possible symmetry which allows for their complete solvability. The modern theory of Superintegrability was inaugurated in 1965~\cite{S:1965,S:1966,S:1967} and was developed further throughout the decades~\cite{C:1969, C:1975, C:2008, C:1990, E:1989, E:2008}. For a thorough review of Superintegrability, we refer to~\cite{T:2004}. In this present work, we will investigate Superintegrability from a Carter-like separability approach of the Hamiltonian.

This paper is organized as follows. In Section \ref{sec:thm}, we generalize Carter's theorem to systems having more degrees of freedom and show that we can obtain non-trivial constants in higher dimensional systems as well. We will prove this by means of the Principle of Mathematical induction. In Section \ref{sec:supint}, we explore the idea of Superintegrability using a `Carter-like' idea and show that Superintegrability can be explored via Carter approach. Finally, in Section \ref{sec:disc}, we summarize with a brief concluding remark. 
%%%%%%%%%%%%%%%%%%%%%%%%%%%%%%%%%%%%%%%%%%%%%%%%
\section{Generalization of Carter's Theorem }\label{sec:thm}
%%%%%%%%%%%%%%%%%%%%%%%%%%%%%%%%%%%%%%%%%%%%%%%%
In this section, we first extend  the Carter's theorem for a system with 3-degrees of freedom, which can be easily generalized to the case of n-degrees of freedom. 
\begin{theorem}\label{th:1}
If a time independent Hamiltonian can be written in the  form : 
\begin{equation}
H = \frac{1}{2} (\frac{H_{1} + H_{2} + H_{3}}{U_{1} + U_{2} + U_{3}})\,,
\label{eq:H3}
\end{equation}
where $U_{1}$, $U_{2}$ and $U_{3}$ are three functions of only $x^1$, $x^2$ and $x^3$ respectively and $H_{1}$ is independent of $p_{2}$,  $p_{3}$ and of all other coordinate functions other than $x^1$,  and similarly for $H_{2}$ and $H_{3}$, then there are two non-trivial  conserved quantities, which  are given by: 
\begin{eqnarray}
\kappa_{1} = & 2 U_1 H- H_1 = & \frac{U_{1} H_{2} + U_{1} H_{3} - U_{2} H_{1}-U_{3} H_{1}}{U_{1} + U_{2} + U_{3}}\, ,
\label{eq:K31} \\
\kappa_{2} = & 2 U_2H - H_2=  &\frac{U_{2} H_{3} + U_{2} H_{1} - U_{1} H_{2}-U_{3} H_{2}}{U_{1} + U_{2} + U_{3}}\,,
\label{eq:K32}
\end{eqnarray}
\end{theorem}
{\it proof:}
We start with the commutation relation:
\begin{equation}
[H_{1},H]  = \frac{1}{2} \left[ H_{1},\frac{H_{1} + H_{2} + H_{3}}{U_{1} + U_{2} + U_{3}}\right]= \frac{1}{2} (H_1 + H_{2} + H_{3}) \left[H_1 , \frac{1}{U_{1}+U_{2}+U_{3}} \right],
\label{eq:p1}
\end{equation}
This is because of the constraints in the form of $H$ and $H_1$ as stated in the theorem, we further have the relation,
\begin{equation}
[U_{1},H]  = \frac{1}{2(U_1 + U_{2} + U_{3})} \left[ U_1, H_1 \right],
\label{eq:p2}
\end{equation}
and,
\begin{equation}
[H_{1},\frac{1}{U_1 + U_{2}+ U_{3}}]  = \frac{1}{(U_1 + U_{2} + U_{3})^2} \left[ U_1, H_1 \right]\,.
\label{eq:p3}
\end{equation}
%%
%%\begin{equation}
%%[H_{1},H]  = \frac{H_1 + H_{2} + H_{3}}{2(U_1 + U_{2} + U_{3})^2} \left[ U_1, H_1 \right],
%%\label{eq:p4}
%%\end{equation}
Combining the above equations, we get:
\begin{equation}
[H_{1},H]  = 2 H[U_1,H],
\label{eq:p5}
\end{equation}
With the above result, now we show that,
\begin{equation}
\kappa_{1} = 2 U_1 H - H_1\,,
\label{eq:p6}
\end{equation}
commutes with the Hamiltonian
\begin{eqnarray}
[H,\kappa_{1}]  &  = [2 U_1 H - H_1, H] =  & 2[U_1 H,H] - [H_1,H]\, \nonumber \\
 =  & 2 H[U_1,H] - [H_1,H] = &  [H_1,H] - [H_1,H] = 0,
\label{eq:p7}
\end{eqnarray}
Here, we observe that $\kappa_{1}$ commutes with $H$ and hence it is a constant of motion. Expanding $\kappa_{1}$, we get
\begin{equation}
\kappa_{1} = 2U_1H-H_1=\frac{U_{1} H_{2} + U_{1} H_{3} - U_{2} H_{1}-U_{3} H_{1}}{U_{1} + U_{2} + U_{3}}\,.
\label{eq:p8}
\end{equation}
Similarly, we can show that $\kappa_2$ is also a constant of motion. It can be checked that $\kappa_1$, $\kappa_2$ and $H$ are functionally independent. Hence, they will contribute to three separate integrals of motion. For  a Hamiltonian, which has the separable form that is given in expression (\ref{eq:H3}),   we see that we can have two more constants of motion $\kappa_2$ and  $\kappa_3$ corresponding to the canonical coordinates $x^2$ and $x^3$ giving two integrals of motion. But the third integral $\kappa_{3}$ is  not  independent,  because the sum of $\kappa_1$, $\kappa_2$ and $\kappa_3$ is zero which is just a constant number. Hence, we  obtain only two independent conserved quantities along with the Hamiltonian using this formalism.
 Next we further generalize this theorem for a system with $n$ degrees of freedom. We extend the theorem to n-dimensions using the Principle of Mathematical Induction, we will get non-trivial constants of motion. 
\begin{theorem}\label{th:2}
If a given Hamiltonian has the form : 
\begin{equation}
H = \frac{1}{2} (\frac{H_{1} + H_{2} + H_{3} +.....+ H_{n}}{U_{1} + U_{2} + U_{3}+.....+U_n})\,,
\label{eq:Hn}
\end{equation}
where $U_{1},  \ U_{2}, \ , \cdots, \ U_{n}$ are n functions of single canonical variable only in $x^1, \ x^2, \ \cdots, \ x^n$ respectively and $H_{1}$ is independent of canonical momentum $p_{2}, \  p_{3}, \ \cdots, \ p_{n}$ and of all other coordinate functions other than $x^1$ and  similar conditions  also hold for functions  $H_{2}, \ H_{3}, \  \cdots,  \ H_{n}$, then there are n-1 independent contants of motions, which are given by:
\begin{eqnarray}
\kappa_{1} & = & \frac{U_{1} H_{2} + U_{1} H_{3} + ....+ U_{1} H_{n} - U_{2} H_{1}-U_{3} H_{1}-....-U_{n} H_{1}}{U_{1} + U_{2} + U_{3}+....+ U_{n}}\,, \nonumber \\
\kappa_{2} &= & \frac{U_{2} H_{1} + U_{2} H_{3} + ....+ U_{2} H_{n} - U_{1} H_{2}-U_{3} H_{2}-....-U_{n} H_{2}}{U_{1} + U_{2} + U_{3}+....+ U_{n}}\,, \nonumber \\
&\vdots & \nonumber \\
\kappa_{n-1} & = & \frac{U_{n-1} H_{1} + U_{n-1} H_{3} + ....+ U_{n-1} H_{n} - U_{1} H_{n-1}-U_{3} H_{n-1}-....-U_{n} H_{n-1}}{U_{1} + U_{2} + U_{3}+....+ U_{n}}\,. \label{eq:nconst}
\end{eqnarray}
\end{theorem}
Expression (\ref{eq:nconst}), along with $H$ gives a set of $n$ conserved quantities.  
From the above theorems, we can infer that even if we are able to separate one coordinate out of all the coordinates, we can obtain a conserved quantity. We state it in a formal manner below:
\begin{theorem}\label{th:3}
If the Hamiltonian has the form :
\begin{equation}
H = \frac{1}{2} (\frac{H_{1} + H_{23}}{U_{1} + U_{23}})\label{eq:H23}
\end{equation}
where,  $U_{1}$ is a function of  canonical variable $x_1$ only.  In addition,  $U_{23}$ is a function of the canonical coordinates $x_2$ and $x_3$ only.  The function $H_{1}$ is independent of $p_{2}, p_3$ and of all coordinates except $x_1$.   While, $H_{23}$ is independent of $p_1$ and of all other coordinate functions other than $x_2$ and $x_3$. If the above conditions are satisfied, the conserved quantity is given by: 
\begin{equation}
\kappa = \frac{U_{1} H_{23} - U_{23} H_{1}}{U_{1} + U_{23}}
\end{equation}
\end{theorem}
For a given coordinate system and Hamiltonian these theorems are useful for  
finding out non-trivial conserved quantities.
It is clear from the above theorems that non-trivial conserved quantities might arise if the Hamiltonian satisfies certain symmetric structures. It should be noted that if $H$ is replaced by any constant of motion, {\it i.e.}  a quantity that commutes with $H$, a similar proof holds. We will explore this idea to observe that Superintegrability can be explained using this simple Carter-like approach. In the next few sections, we will see the implications of Carter constant to the ideas of superintegrability and also construct some examples of Carter's constant in specific Hamiltonians. 
%%%%%%%%%%%%%%%%%%%%%%%%%%%%%%%%%%%%%%%%%%
%\section{ Applications of Carter Theorem}
%
%We will now propose a new perspective to explain Superintegrability using Carter method and will see some examples of Carter Constant which allows conserved quantities for dipole-like potentials in Newtonian dynamics. We will also give some General Relativistic examples for Carter Constant, though the physical significance of such metrics will not be realized in this paper. 
 
\section{Application to Superintegrable Hamiltonians \label{sec:supint}}
We can use Carter Constant approach to identify all the conserved quantities in several Superintegrable systems as it is based on the observation of separability of the Hamiltonan.  An n-dimensional system is Superintegrable if there exists more than \textit{n} functionally independent globally defined and single valued integrals.  We will explore Superintegrability in 2- and 3- dimensional systems by constructing explicit examples of Superintegrable systems using Carter criteria. For Hamiltonian systems with three degrees of freedom,  Evans~\cite{E:1989} carried out a detailed investigation and tabulated the conserved quantities in various possible separable coordinate systems. In order that our examples don't overlap with Evans we will construct superintegrable examples in 2-D systems using Carter separability criteria. Because Evans had already classified all 3-D superintegrable systems thoroughly, at the end we would show that all 3-D classical superintegrable systems are also Carter separable. That would be to provide an instructive aspect to our approach. Since we are considering Hamiltonians which does not explicitly depend on time, the Hamiltonian or the energy ($E$) itself is one of the integrals of motion. We would construct three different Carter separable and Superintegrable systems by constructing Hamiltonians that are separable in more than one coordinate systems. These three examples would be for illustrative purposes.\\
\textbf{Example 1}

We start with the following Hamiltonian in 2-D Cartesian coordinate system:
\begin{equation}
H= \frac{1}{2}({p_x^2 + p_y^2 }) + \frac{1}{(x^2+y^2)^{1/2}}\left(\alpha+\frac{\beta}{(x+\sqrt{x^2+y^2})}+ \frac{\gamma}{(\sqrt{x^2+y^2}-x)} \right) \,.
\end{equation}

We will show that this system is superintegrable by bringing it to Carter separable form in polar and parabolic cylindrical coordinates. In polar coordinates, $x=r~cos\theta$, $y=r~sin\theta$, and Hamiltoninan (17) becomes:
\begin{equation}
H= \frac{1}{2}({p_r^2 + \frac{p_{\theta}^2}{r^2} }) + \frac{1}{r}\left(\alpha+\frac{\beta}{(r~cos\theta+ r)}+ \frac{\gamma}{(r-r~cos\theta)} \right) \,.
\end{equation}
simplified to:
\begin{equation}
H= \frac{1}{~r^2}\left(\frac{(r~p_r)^2}{2} + \frac{p_{\theta}^2}{2}  + \alpha~r+\frac{\beta}{(~cos\theta+ 1)}+ \frac{\gamma}{(1-~cos\theta)} \right) \,.
\end{equation}

In this form, it is clear that $r$ and $\theta$ components separate, and a constant of motion is $K_1 = \frac{p_{\theta}^2}{2}+\frac{\beta}{(~cos\theta+ 1)}+ \frac{\gamma}{(1-~cos\theta)}$. A second constant of motion is trivially the Hamiltonian, $K_2=H$. For the third integral of motion, we look at cylindrical parabolic coordinates. Parabolic cylindrical coordinates are defined by $\xi = \sqrt{x^2+y^2}+x$ and $\eta = \sqrt{x^2+y^2}-x$, and the Hamiltonian becomes:
\begin{equation}
H= \frac{\xi^2~p_{\xi}^2 + \eta^2~p_{\eta}^2+ \alpha + \frac{\beta}{\xi} + \frac{\gamma}{\eta}}{\eta + \xi} \,.
\end{equation}
In this form, it is easy to see that the Hamiltonian is Carter separable, with $H_{\xi}= \xi^2~p_{\xi}^2+ \frac{\beta}{\xi}$, $H_{\eta}= \eta^2~p_{\eta}^2+ \frac{\gamma}{\eta}$, $U_{\xi}=\xi$ and $U_{\eta}=\eta$, and so we have a superintegrable system with the extra integral of motion being:
\begin{equation}
K_3 = \frac{\xi(\eta^2 p_{\eta}^2 + \frac{\gamma}{\eta})-\eta(\xi^2 p_{\xi}^2 + \frac{\beta}{\xi})}{\eta+\xi} \,.
\end{equation}

Hence, this system is superintegrable as can be seen from Carter separability.

\textbf{Example 2}
We start with the following Hamiltonian in 2-D Cartesian coordinate system:
\begin{equation}
H= \frac{1}{2}({p_x^2 + p_y^2 }) + \frac{1}{\sqrt{x^2+y^2}}\left(\alpha+\beta\sqrt{x+\sqrt{x^2+y^2}}+ \gamma\sqrt{\sqrt{x^2+y^2}-x} \right) \,.
\end{equation}

In rotational parabolic coordinates, $x=\sigma\tau$ and $y=\frac{1}{2}(\tau^2-\sigma^2)$ and $x^2+y^2=\frac{1}{4}(\sigma^2+\tau^2)^2$, and the momenta squared is found out from the Laplacian to be: $p_x^2+p_y^2=\frac{1}{\sigma^2+\tau^2}[p_{\sigma}^2+p_{\tau}^2]$. Expanding the Hamiltonian (22), in this coordinate system gives: 
\begin{equation}
H= \frac{1}{\sigma^2+\tau^2}\left[p_{\sigma}^2+p_{\tau}^2+\alpha+\beta(\sigma+\tau)+\gamma(\sigma-\tau)\right] \\ = \frac{1}{\sigma^2+\tau^2}\left[p_{\sigma}^2+p_{\tau}^2+\alpha+(\beta+\gamma)\sigma+(\beta-\gamma)\tau\right]\,.
\end{equation}
In this form, clearly this Hamiltonian is Carter separable in rotational parabolic coordinates, with the corresponding Carter constant given by:
\begin{equation}
K_1= \frac{1}{\sigma^2+\tau^2}\left[\tau^2[p_{\sigma}^2+(\beta+\gamma)\sigma]-\sigma^2[p_{\tau}^2+(\beta-\gamma)\tau]\right]\,.
\end{equation}
The second constant of motion is the Hamiltonian itself ($K_2=H$), and the third extra integral, we see that we can separate this system in parabolic cylindrical coordinates, where the Hamiltonian is:
\begin{equation}
H= \frac{1}{\eta+\xi}\left[\eta^2p_{\eta}^2+\xi^2p_{\xi}^2+\alpha+\beta~\sqrt{\xi}+\gamma~\sqrt{\eta}\right]\,.
\end{equation}
Clearly, giving Carter separability again and another constant of motion $K_3=\frac{1}{\eta+\xi}[\xi(\eta^2p_{\eta}^2+\gamma\sqrt{\eta})-\eta(\xi^2p_{\xi}^2+\beta\sqrt{\xi})]$, so, with $K_1,K_2$ and $K_3$, this 2-D system becomes Superintegrable. \\
\textbf{Example 3}
We start with the following Hamiltonian in 2-D Rotational parabolic system:
\begin{equation}
H= \frac{1}{\sigma^2 + \tau^2}({p_{\sigma}^2 + p_{\tau}^2 }) + \tau^2 - \sigma^2 \,.
\end{equation}
which can be also written as:
\begin{equation}
H= \frac{1}{\sigma^2 + \tau^2}({p_{\sigma}^2 + p_{\tau}^2 } + \tau^4 - \sigma^4) \,.
\end{equation}
having the form $\frac{H_{\sigma}+H_{\tau}}{U_{\sigma}+U_{\tau}}$, with $H_{\sigma}=p_{\sigma}^2- \sigma^4$, $H_{\tau}=p_{\tau}^2+ \tau^4$, $U_{\sigma}=\sigma^2$ and $U_{\tau}=\tau^2$, giving the conserved quantity, $K_1= \frac{\sigma^2(p_{\tau}^2+\tau^4)- \tau^2(p_{\sigma}^2-\sigma^4)}{\sigma^2+\tau^2}$, in Cartesian coordinate system, this Hamiltonian takes on the simple form $H=p_{x}^2 + p_{y}^2 + 2y$, thereby giving a 2nd integral $K_2=p_x$, and along with $K_3=H$, we see that this Hamiltonian admits 3 integrals becoming Superintegrable.

The linear independence of these conserved quantities is an easy check. Many such examples of Superintegrable systems can be constructed in 2-dimensions by following the general idea of constructing a Carter separable Hamiltonian in two different coordinate systems, i.e. we look for a Hamiltonian of the form $H=\frac{1}{2}\left(\frac{H_r+H_{\theta}}{U_r+U_{\theta}}\right)$ in some coordinate system $(r,\theta)$ and then we also demand that this same Hamiltonian has the form $H=\frac{1}{2}\left(\frac{H_{\eta}+H_{\xi}}{U_{\xi}+U_{\eta}}\right)$ in some different coordinate system $(\eta,\xi)$, then along with the conserved quantity $H$, we will also have $K_1= \frac{U_r H_{\theta}-U_{\theta}H_r}{U_r+U_{\theta}}$ and $K_2= \frac{U_{\eta} H_{\xi}-U_{\xi}H_{\eta}}{U_{\eta}+U_{\xi}}$ as other conserved quantities hence, if we are then able to prove the linear independence of these constants, we will be able to claim that the corresponding system is Superintegrable. This general procedure can then be carried over to systems with higher degrees of freedom as well in accordance with the theorems presented in Section 2. Of course, it must be kept in mind that this idea may not be able to reproduce all possible Superintegrable conserved quantities for all possible system howsoever complicated it might be. But, it is clear that this provides an alternative approach to construct and explore at simple and non-trivial Superintegrable systems. 
 
For 3-dimensional systems, Evans paper ~\cite{E:1989} gives an in-depth analysis of all possible Superintegrable systems. Apart from the 2-D examples that we constructed above, we will take one sample potential from Evan's paper and show that all the potentials listed there are also Carter separable.
\begin{equation}
H= p_{x}^2 + p_{y}^2+ p_{z}^2-\frac{k}{r} + \frac{k_1}{x^2}+ \frac{k_2}{y^2} \,.
\end{equation}
We look at the Hamiltonian in spherical polar coordinates, which takes the form:
\begin{equation}
H= \frac{1}{2}(p_r^2 + \frac {p_{\theta}^2}{r^2} + \frac{ p_{\phi}^2}{r^2\sin^2{\theta}} )-\frac{k}{r} + \frac{k_1}{r^2 \sin^2{\theta} \cos^2{\phi}}+ \frac{k_2}{r^2 \sin^2{\theta} \sin^2{\phi}}\,.
\label{eq:s2}
\end{equation}
Here, we can separate  the $r$ coordinate from  $H$, then apply \textit{theorem~\ref{th:3}} to obtain the following constant:
\begin{equation}
I_1 = \frac{1}{2} (p_{\theta}^2 + \frac{p_{\phi}^2}{\sin^2{\theta}}) + \frac{k_1}{\sin^2{\theta} \cos^2{\phi}} + \frac{k_2}{\sin^2{\theta} \sin^2{\phi}}\,.
\label{eq:s3}
\end{equation}
It can be seen that $I_1$ can be further separated by taking out $\sin^2{\theta}$ as a common factor. Since  $I_1$ commutes with $H$, we can further apply the \textit{theorem~\ref{th:3}} on $I_1$, resulting in a  second integral of motion that reads:
\begin{equation}
I_2 = \frac{1}{2} p_{\phi}^2 + \frac{k_1}{\cos^2{\phi}} + \frac{k_2}{\sin^2{\phi}}\,.
\label{eq:s4}
\end{equation}
Along with $I_1$, $I_2$ and total energy ($E$), the  system has three conserved quantities and is integrable.   For system to be superintegrable, we need at least one more independent conserved quantity. Now we use another coordinate system in which the system is separable, {\it i.e. } rotational 
parabolic coordinate system, $(\xi, \, \eta, \, \phi)$ where the coordinate transformation are given by:
\begin{eqnarray}
x=\xi\,\eta\, \cos{\phi},  &
y=\xi\,\eta\, \cos{\phi},  &
z=\frac{1}{2} (\eta^2 - \xi^2)\,,
\label{eq:s5}
\end{eqnarray}
where $\xi\geq 0, \ \eta<\infty$, and $ 0\leq \phi \leq 2\pi $.  Under this coordinate transformation the Hamiltonian can be written as: 
\begin{equation}
H=  \frac{1}{2(\eta^2 + \xi^2)}\left[ p_{\xi}^2 + p_{\eta}^2 + \frac{(\eta^2+\xi^2)p_{\phi}^2}{\xi^2\eta^2} - 4k + 2k_1 \frac{(\eta^2 + \xi^2)}{\xi^2\eta^2 \cos^2{\phi}} + 2k_2 \frac{(\eta^2 + \xi^2)}{\xi^2\eta^2 \sin^2{\phi}}\right] \,.
\label{eq:s6}
\end{equation}
We now observe that the last two terms and the $p_{\phi}$ in the above expression is nothing but, $(\frac{1}{\eta^2}+ \frac{1}{\xi^2}) I_2$, so the form of the Hamiltonian can be further simplified to:
\begin{equation}
H=  \frac{1}{2(\eta^2 + \xi^2)}\left[ p_{\xi}^2 + p_{\eta}^2 - 4k + (\frac{1}{\eta^2}+ \frac{1}{\xi^2}) I_2 \right]\,.
\label{eq:s7}
\end{equation}
This equation is clearly separable and we can apply \textit{theorem~\ref{th:3}} to  obtain another conserved quantity with $U_{\xi}= \xi^2$ and  $U_{\eta}= \eta^2$, 
\begin{equation}
I_4=\frac{\eta^2(p_{\xi}^2 + \frac{I_2}{\xi^2} - 2k) -\xi^2(p_{\eta}^2 + \frac{I_2}{\eta^2}- 2k)}{\eta^2+\xi^2}\,.
\label{eq:s8}
\end{equation}
Substitutig $I_2$ back, we obtain:
\begin{equation}
I_4=\frac{\eta^2 p_{\xi}^2 - \xi^2 p_{\eta}^2}{\eta^2 + \xi^2}+ (\eta^2 - \xi^2) (\frac{k_1}{\xi\eta \cos^2{\phi}}+ \frac{k_2}{\xi\eta \sin^2{\phi}}- \frac{2 k}{\eta^2 + \xi^2 })\,.
\label{eq:s9}
\end{equation}
With  $I_4$ obtained in this way gives rise to a non-trivial and functionally independent conserved quantity of the Hamiltonian thereby giving us four possible conserved quantities of motion, thus making the system Superintegrable. On doing this exercise for all potentials listed in ~\cite{E:1989}, we will be able to see that all these 3-D superintegrable potentials are also Carter separable in more than one coordinate systems. 

So, to summarize, in this section we have given a simple approach of probing certain Superintegrable systems via Carter approach by direct inspection of the Hamiltonian. We have demonstrated our proposition using three examples of Superintegrable Hamiltonians in 2-dimensions and following a Carter-like approach to perceive the associated Superintegrability. We have also verified that all Superintegrable potentials in 3-Dimensional Newtonian dynamics are also Carter separable. The question of whether this simple idea can be applied to extremely complicated Hamiltonian is indeed left open for now and as a scope for future work.
%%%%%%%%%%%%%%%%%%%%%%%%%%%%%%%%%%%%%%%
%%%%%%%%%%%%%%%%%%%%%%%%%%%%%%%%%%%%%%%

\section{discussions} \label{sec:disc}
In this work, we explored the hidden symmetries of Superintegrable systems by means of the presence of Carter constant as an integral of motion. We constructed some 2-Dimensional superintegrable systems by constructing potentials that are Carter separable in more than one coordinate systems. Many such 2-D Superintegrable potentials may be constructed by following the general procedure outlined in Section 3. We constructed 2-D non-trivial Superintegrable systems because all possible 3-D superintegrable sytems has been outlined by Evans in ~\cite{E:1989}. But we have verified that all of those potentials in 3-D also satisfy this Carter Separability criteria. This general construction of simple non-trivial Superintegrable systems can be extended to higher dimensions by algebraic construction of Carter separable systems in more than one coordinate systems and and in turn checking the linear independence of these various Carter-like conserved quantities arising out of them.

\section{Acknowledgements}
K. R. N wish to thank the Visiting Associateship programme of Inter-University Centre for Astronomy and Astrophysics (IUCAA), Pune. A part of this work was carried out during the visit to IUCAA under this programme. Many thanks to Dr. Ananda Dasgupta for useful questions.

\begingroup

\begin{thebibliography}{apacite}
\bibliographystyle{abbrv}

%\cite{R:2003abc}
\bibitem{Carter:2003abc} 
B. Carter,
Phys. Rev, 174 , 1559 (1968).

\bibitem{M:1973} 
C.W. Misner, K.S. Thorne, and J.A.Wheeler, Gravitation (W.H. Freeman, San Francisco, 1973).

%\cite{C:1973abc}
\bibitem{C:1973abc}
B. Carter, Ed. De Witt and B. S. De Witt, Gordon and Breach, New York (1973).

%\cite{SM:2015abc}
\bibitem{Sajal:2015abc}
   S.~Mukherjee and K.~R.~Nayak
  %``Carter constant and angular momentum ,''
  (arxiv: 1507:01863) (2015).
  
%\cite{W:2009abc}
\bibitem{Will:2009abc} 
  C. M. Will, Phys. Rev. Lett, 102, 061101, (2009).
  
%\cite{K:20077abc}
\bibitem{Rosquist:20077abc}
   K.~Rosquist, T.~Bylund, L.~Samuelsson,
  %``Carter's constant revealed ,''
  Int. J. Mod. Phys, D18, 429-434 (2009). 
  
%\cite{R:34abc}
\bibitem{Fel:1999abc}  
	F. de.~ Felice and G. Preti,
	Class. Quantum Grav, 16 , 2929 (1999). 

\bibitem{Wa:1970abc}   
  M. Walker and R. Penrose, Commun. Math. Phys, 18 , 265 (1970).
  
%\cite{R:2003abc}
\bibitem{Ramachandra:2003abc}
   B. S.~Ramachandra,
  %``The Carter Constant and the Petrov Classification of the VEK Spacetime ,''
 ``The Carter Constant and the Petrov Classification of the VEK Spacetime" , Indian Institute of  	   Astrophysics, University of Calicut (2003).
 
\bibitem{K:1963}
R. P. Kerr, Phys. Rev. Lett, 11 , 237 (1963).

\bibitem{S:1965}
I. Fris , V. Mandrosov , J. A. Smorodinsky, M. Uhlir and P. Winternitz, Phys. Lett, 16, 354–356 (1965).

\bibitem{S:1966}
I. Fris , J. A. Smorodinsky, M. Uhlir and P. Winternitz, Yad Fiz, 4, 625–635 (1966).

\bibitem{S:1967}
A. A. Makarov, J. A. Smorodinsky, K. Valiev and P. Winternitz, Il Nuovo Cimento A, 52, 1061–1084 (1967).

\bibitem{C:1969}
F. Calogero, J. Math. Phys, 10, 2191–2196 (1969).

\bibitem{C:1975}
F. Calogero, Lettere al Nuovo Cimento, 13, 411–416 (1975).

\bibitem{C:2008}
C. Chanu, L. Degiovanni and G. Rastelli, J. Math. Phys, 49, 112901 (2008).

\bibitem{C:1990}
N. W. Evans, Phys. Lett, 147, 483–486 (1990).

\bibitem{E:1989}
N. W. Evans , Phys. Rev A, 41, 10, 1989.

\bibitem{E:2008}
N. W. Evans and P. E. Verrier, J. Math. Phys, 49, 092902 (2008).

\bibitem{T:2004}
P. Tempesta, P. Winternitz, W. Miller and G. Pogosyan, CRM Proceedings and Lecture Notes, 37 (2004).


\end{thebibliography}
\endgroup
\end{document}